# High expression of antioxidant proteins in dendritic cells : possible implications in atherosclerosis

Aymeric Rivollier[1], Laure Perrin-Cocon[1], Sylvie Luche[2], Hélène Diemer[3], Jean-Marc Strub[3], Daniel Hanau[4], Alain van Dorsselaer[3], Vincent Lotteau[1], Chantal Rabourdin-Combe[1] ,Thierry Rabilloud[2] and Christine Servet-Delprat[1]

[1] INSERM U503, Université Lyon 1, IFR128-Biosciences Gerland, 21 avenue Tony Garnier, 69 007 LYON cedex 07, France
[2] CEA, Laboratoire d'Immunochimie, DRDC/ICH, INSERM U 548, CEA-Grenoble, 17 rue des martyrs, F-38054 GRENOBLE CEDEX 9, France
[3] Laboratoire de Spectrométrie de Masse Bio-Organique, UMR CNRS 7509, ECPM, 25 rue 3 Becquerel, 67008 STRASBOURG Cedex, France
[4] Biologie des Cellules Dendritiques Humaines, INSERM U725, ETS Strasbourg, 10 rue Spielmann , 67 065 Strasbourg cedex, France

Correspondence :
Thierry Rabilloud, DRDC/ICH, INSERM U 548
CEA-Grenoble, 17 rue des martyrs,
F-38054 GRENOBLE CEDEX 9
Tel (33)-4-38-78-32-12
Fax (33)-4-38-78-51-87
e-mail: Thierry.Rabilloud@cea.fr

Abbreviations:
DC, dendritic cells, LPS: lipopolysaccharide , MHC: major histocompatibility complex, MnSOD: manganese superoxide dismutase, OxLDL: oxidized low-density lipoprotein , Prx: peroxiredoxin

Abstract
Dendritic cells (DC) displays the unique ability to activate naive T cells and to initiate primary T-cell responses revealed in DC – T cell alloreactions. DC frequently operate under stress conditions. Oxidative stress enhances the production of inflammatory cytokines by DC. We performed a proteomic analysis in order to see which major changes occur, at the protein expression level, during DC differentiation and maturation. Comparative 2D gel analysis of the monocyte, immature DC and mature DC stages were performed. Manganese superoxide dismutase (MnSOD), reaches 0.7% of the gel-displayed proteins at the mature DC stage. This important amount of MnSOD is a primary antioxidant defence system against superoxide radicals, but its product, $H_2O_2$, is also deleterious for cells. Peroxiredoxin enzymes (Prx) play an important role in eliminating such peroxide. Prx1 expression level continuously increases during DC differentiation and maturation while Prx6 continuously decreases and Prx2 peaks at the immature DC stage. As a consequence, DC are more resistant than monocytes to apoptosis induced by high amounts of oxidized low-density lipoproteins (oxLDL) containing toxic organic peroxides and hydrogen peroxide. Furthermore, DC-stimulated T cells produce high levels of receptor activator of nuclear factor B ligand (RANKL), a chemotactic and survival factor for moncocytes and DC. This study provides insights into the original ability of DC to express very high levels of antioxidant enzymes such as MnSOD and Prx1, to detoxify oxLDL and to induce high levels of RANKL by the T cells they activate, and further emphasizes the role that DC might play in atherosclerosis, a pathology recognized as a chronic inflammatory disorder.

**1-Introduction**

Dendritic cells (DC) are the most potent antigen-presenting cells (APC) in the body, and their unique ability to stimulate a primary T-cell response places them at the centre of the immune response [1]. Immature DC differentiate from bone marrow progenitors or from monocytes and then either stay in the blood stream or migrate in the peripheral tissues. Immature DC, such as Langerhans cells in the skin, survey incoming pathogens. They are equipped with receptors to become activated when exposed to pathogen associated molecular patterns (PAMP) [2]. Their capacity to recognize pathogens and become activated therefore represents the first critical event in the initiation of the immune response. Encounter with a pathogen leads to DC maturation and migration through lymphatic vessels to T cell area of secondary lymphoid organs. Antigen presentation by DC activates specific naive T cells to express CD40 ligand (CD154) [3], which, in turn, activates DC, achieving their terminal differentiation, as assessed by the up-regulation of MHC-I and II molecules and of co-stimulatory molecules CD80/CD86 and by the production of cytokines, such as IL-12 and IL-1$\alpha$/$\beta$, which all participate in T cell stimulation [3-5] and in the development of adaptive immunity [4], [6]. Mature DC modulate T-cell responses through the secretion of various cytokines such as IL-12, promoting Th1-type cellular immune response [7] or IL-4 following thymic stromal lymphopoietin activation, promoting Th2-type humoral immune response [8]. Last, a DC apoptosis program can be triggered at the end of the maturation process, so that mature DC do not produce an over stimulation of the immune system [9]. DC frequently operate under stress conditions induced by tissue damage, infectious pathogens or inflammatory reactions. Oxidative stress enhances the production of inflammatory cytokines by DC [10].

Atherosclerosis is considered as a chronic inflammatory disease of the arterial vasculature, initiated by endothelial cell damage and implicating both monocytes, DC and macrophages that operate under intense oxidative stress conditions. Indeed, the accumulation of oxidized low-density lipoproteins (oxLDL), generated from native LDL trapped into the subendothelial space of the arterial wall, is a main feature of the disease and plays a key role in its progression, leading to the formation of vascular lesions [11], [12]. This accumulation of oxLDL induces the activation of macrophages and elicits an oxidative burst [13], generating toxic components such as the superoxide anions, derived from the NADPH oxidase activity. Lipid peroxides and hydrogen peroxide directly contained in oxLDL are also toxic per se [14]. Due to their pro-oxidant actions, superoxide anions resulting from macrophage activation by oxLDL, as well as lipid peroxides contained in the oxLDL, are able to induce resident monocytes and macrophages apoptosis [14-16]. This apoptosis is a commun feature within early atherosclerotic lesions [17]. Superoxide anions can be reduced into $H_2O$ and $O_2$ by combination of the actions of cellular superoxide dismutase (SOD) and peroxiredoxin (Prx) or catalase or other peroxidases. Thus, it is hypothesized that monocytes and macrophages undergo apoptosis when their antioxidant enzymatic protection systems (SOD and Prx) are overflowed. Contrary to monocytes and macrophages, the expression pattern of oxidative stress response proteins in DC is still unknown. Although both immature and mature DC have been observed in the atherosclerotic plaques in close association with activated T cells [18], and in para-aortic and jugulodigastric lymph nodes attached to atherosclerotic arterial wall [19], their exact role in atherosclerosis progression is still poorly understood. These studies however suggest that vascular DC may be implicated in the local induction of immune and auto-immune reactions [20], [21].

Besides their toxic action on diverse cell types, OxLDL also increase monocyte adhesion to the endothelial cell layer, as well as their transmigration [22-24] towards adjacent tissues. Finally oxLDL trigger the production of pro-inflammatory cytokines such as monocyte chemoattractant protein-1 (MCP-1), M-CSF, and granulocyte macrophage colony-stimulating factor (GM-CSF) by endothelial cells and stimulate monocyte differentiation into DC [25].

Here we investigated the relationship between oxidative stress protein expression profiles and DC differentiation and maturation. We performed a proteomic analysis of both immature and mature DC derived from monocytes in the presence of IL-4 and GM-CSF and compared it to monocyte proteome. This comparative proteomic approach revealed the original ability of DC to express high levels of antioxidant enzymes such as MnSOD and Prx. These high expression levels of antioxydant enzymes confer a great capacity to DC to resist to apoptosis induced by oxLDL, mimicking the oxidative stress microenvironment of atherosclerosis lesions. Moreover, when co-cultured with T cells, DC produce themselves, and induce T cells to produce, high levels of receptor activator of NF-κB ligand (RANKL), a chemotactic factor for monocytes and a key survival factor for DC [26-28]. Altogether, these *in vitro* results suggest that DC play a crucial role in the atherosclerosis pathogenesis and in the maintenance of its chronicity.

## 2-Materials and methods

### Cell culture and reagents

Monocytes and T cells were purified [29] from the adult blood of healthy volunteer donors (Etablissement français du sang, Lyon Gerland, France). Monocyte-derived DC were generated in vitro, as previously described [29]. Briefly, monocytes were seeded at $10^6$ cells/mL and maintained in RPMI 1640 (Life Technologies, Paisley, Scotland) supplemented with 10 mM Hepes, 2 mM L-glutamine, 40 µg/mL gentamicin (Life Technologies), 10% heat-inactivated FCS (Boehringer Mannheim, Meylan, France), 50 ng/mL hrGM-CSF and 500 U/mL hrIL-4. After 6 days of culture, more than 95% of the cells were DC as assessed by CD1a labelling. Recombinant human GM-CSF and IL-4 were purchased from PeproTech (Rocky Hill, New Jersey). For maturation, immature DC were plated at $10^6$ cells/ml and were stimulated with 1 µg/mL LPS or with $10^5$/ml irradiated (7,000 rads) fibroblastic CD40L- or control CD32- transfected L cells (both kindly provided by Schering-Plough Laboratory for Immunological Research, Dardilly, France) for 24h. For OxLDL cultures, 10% lipoprotein-deficient FCS (Sigma-Aldrich) was used. T cell activation was performed at $10^6$ cells/mL with 1µg/mL anti-CD3 (HIT3a murine mAb) and 10µg/mL anti-CD28 (CD28.2 murine mAb) from PharMingen (California). For proteomic studies, cells were harvested by centrifugation, rinsed in PBS and resuspended in homogeneization buffer (0.25 M sucrose, 10 mM Tris-HCl, pH 7.5, 1 mM EDTA). A buffer volume approximately equal to the packed cell volume was used. The suspension was transferred to a polyallomer ultracentrifuge tube and the cells were lysed by the addition of 4 volumes (respective to the suspension volume) of 8.75 M urea, 2.5 M thiourea, 25 mM spermine base and 50 mM DTT. After 1 hour at room temperature, the extracts were ultracentrifuged (30 minutes at 200000g). The supernatant was collected and the protein was determined by a Bradford assay, using bovine serum albumin as a standard. Carrier ampholytes (0.4 % final concentration) were added and the protein extracts were stored at -20°C.
Human Prx2 was purified from freshly isolated human red blood cell by ion-exchange chromatography and gel filtration, as previously described [30].

### LDL preparation

LDL ($1.025 \leq$ density $\leq 1.055$ g/ml) was isolated from human plasma of normolipidemic healthy individuals by ultracentrifugation, as described previously [25]. The protein content of LDL was estimated by Coomassie Protein MicroAssay procedure (Pierce, Rockford, IL), and its lipid composition was determined using cholesterol RTU, triglycerides enzymatic PAP 150, and phospholipids enzymatic PAP 150 kits from bioMérieux (Marcy l'Etoile, France).

### LDL oxidation

LDL concentration was adjusted at 1 mg/ml of protein by dilution in PBS and dialyzed at 4°C against PBS to eliminate EDTA. $Cu^{2+}$-mediated oxidation was conducted at 37°C for 24 h by dialysis against 5 µM CuSO4/ PBS. The reaction was stopped by addition of 40 µM butylated-hydroxytoluene and extensive dialysis at 4°C against PBS containing 100 µM diethylenediamine pentaacetic acid.

**Flow cytometry.**
Cell suspensions were labelled according to standard procedures using the following monoclonal antibodies: CD1a-PE, CD14-PE, CD25-PE, CD80-FITC, CD83-PE, CD86-PE, MHC I (HLA-ABC-FITC), MHC II (HLA-DR-FITC) or an isotype control (Beckman Coulter, Villepinte, France). 0.5 µg/mL of propidium iodide were added to detect apoptotic cells. Immunostaining was performed in 1% BSA and 3% human serum-PBS and then quantified on a FACSCalibur (Becton Dickinson, Pont de Claix, France).

**Immunofluorescence staining.**

Cells cultured on glass coverslips were first fixed for 10 min with 3.7% formaldehyde in phosphate buffered saline (PBS) and permeabilized with 0.1% Triton X-100 in PBS for 7 min. After pre-incubation for 20 min in normal human serum with 10% PBS, cells were incubated with anti-RANKL (sc-9073 rabbit polyclonal Santa-Cruz) and anti-CD3 (UCHT1 mAb Beckman Coulter) antibodies. Coverslips were then treated with the appropriate conjugated secondary antibodies (donkey anti-rabbit or donkey anti-mouse antibodies, Jackson ImmunoResearch). Primary and secondary antibodies were applied for 60 min, in a humidified chamber. Between each step, coverslips were washed three times for 5 min in PBS buffer. Observations were performed by epifluorescence using a Zeiss axioplan microscope.

**Allogeneic T cell stimulation.**
DC were cultured in various numbers (10 to $10^5$), for 7 days, in the presence of a constant number of T cells ($10^5$ cells / well) purified from the blood of another donor (allogeneic), as previously described [31]. [$^3$H] Thymidine incorporation was measured after a 12h pulse with 1 µCi [$^3$H]TdR/ well, using a Top Count NXT counter (Packard Bioscience, PerkinElmer Life Sciences, France).

**Western blots**
Cells were lysed in a buffer containing 200 mM NaCl, 40 mM Tris–HCl pH 8.0, 1% NP-40 2 mM EDTA, 1mM PMSF, 1mM NaF, 10µg/mL aprotinin and a mixture of protease inhibitors (protease inhibitor set III, Calbiochem, Darmstadt, Germany) for 15 min at 4°C. Insoluble materials were removed by centrifugation at 10 000 g for 10 min. Proteins from cell lysates were separated by SDS–PAGE using NuPAGE 4-12% Bis-Tris Gel (Invitrogen, Cergy Pontoise, France), then transferred to Immobilon-P membranes (Millipore, Bedford, MA). Membranes were blocked using 5% BSA in TBS-T (20 mM Tris (pH 7.6), 130 mM NaCl, and 0.1% Tween 20) and incubated for 1 h with a specific anti-MnSOD antibody (#06-984, Upstate cell signaling solutions, Charlottesville, VA). Immunoreactive bands were visualized by using a secondary goat anti-rabbit HRP-conjugated antibody (Jackson ImmunoResearch, West Grove, PE) and chemiluminescence (ECL Western Blotting Substrate Kit, Pierce, Rockford, IL). The membranes were not stripped before reblotting with anti-actin antibody (A-2066, Sigma-Aldrich Chimie, Saint Quentin Fallavier, France).

**Two-dimensional electrophoresis**
Two-dimensional electrophoresis was performed with immobilised pH gradients for isoelectric focusing. Home-made linear 4-8 or 4-12 gradients were used [32] and prepared according to published procedures [33]. IPG strips were cut with a paper cutter, and rehydrated in 7M urea, 2M thiourea, 4%

CHAPS, 0.4% carrier ampholytes (3-10 range), containing either 20mM DTT (4-8 gradients) or 5mM Tris cyanoethyl phosphine (purchased from MolecularProbes, for 4-12 gradients) [34]. The protein sample was mixed with the rehydration solution in the case of 4-8 gradients, or cup-loaded at the anode for 4-12 gradients. Isoelectric focusing was carried out for a total of 60000 Vh. After focusing, the strips were equilibrated for 2 x 10 minutes in 6M urea, 2% SDS, 125 mM Tris-HCl pH 7.5 containing either 50mM DTT (first equilibration step) or 150mM iodoacetamide (second equilibration step). The equilibrated strip was loaded on the top of a 10% or 11% polyacrylamide gel, and submitted to SDS PAGE at 12W/ gel, using the Tris-taurine system [35].

After migration, the gels were stained either with silver nitrate for 2D gels with a pH 4-8 gradient [36]., or with ammoniacal silver for 2D gels with a pH 4 to 12 gradient [37]. Quantitative gel analysis was performed on the silver-stained gels with the Melanie II software (Genebio, Geneva, Switzerland). The experiments were performed in triplicate, starting with different cell batches. Several gels were made for each culture, in order to select gels with very close detection signal levels for quantitative analysis. This allowed us to keep the gel analysis parameters constant for better reproducibility. As a matter of facts, the total spot intensity in the analyzed gels ranged from 531400 to 662750 arbitrary absorbance units, with a mean of 600290 units, i.e. a maximum deviation of ±11%.

**Mass spectrometry**

*In gel digestion :*

Excised gel slice rinsing was performed by the Massprep (Micromass, Manchester, UK) as described previously [38]. Gel pieces were completely dried with a Speed Vac before digestion. The dried gel volume was evaluated and three volumes trypsin (Promega, Madison, US) 12.5ng/$\mu$l freshly diluted in 25mM $NH_4HCO_3$, were added. The digestion was performed at 35°C overnight. Then, the gel pieces were centrifuged for 5 min in a Speed Vac and 5$\mu$l of 35% $H_2O$/ 60% acetonitrile/ 5% HCOOH were added to extracted peptides. The mixture was sonicated for 5 min and centrifuged for 5 min. The supernatant was recovered and the procedure was repeated once.

*MALDI-TOF-MS analysis*

Mass measurements were carried out on an ULTRAFLEX™ MALDI TOF/TOF mass spectrometer (Bruker-Daltonik GmbH, Bremen, Germany).This instrument was used at a maximum accelerating potential of 20kV and was operated either in reflector positive mode. Sample preparation was performed with the dried droplet method using a mixture of 0.5mL of sample with 0.5mL of matrix solution. The matrix solution was prepared from a saturated solution of α-cyano-4-hydroxycinnamic acid in $H_2O$/ 50% acetonitrile diluted 3 times. Internal calibration was performed with tryptic peptides resulting from autodigestion of trypsin (monoisotopic masses at m/z=842.51 , m/z=1045.564 , m/z= 2211.105).

*MS Data analysis*

Monoisotopic peptide masses were assigned and used for databases searches using the search engines MASCOT (Matrix Science, London, UK) [39], and Aldente (www.expasy.org). All proteins present in Swiss-Prot were used without any pI and Mr restrictions. The peptide mass error was limited to 90 ppm, one possible missed cleavage was accepted.

*MS/MS data analysis*

(LC-MS-MS) analysis of the digested proteins were performed using a CapLC capillary LC system (Micromass, Manchester, UK) coupled to a hybrid quadrupole orthogonal acceleration time-of-flight tandem mass spectrometer (Q-TOF II, Micromass). The LC-MS union was made with a PicoTip (New Objective, Woburn,.MA) fitted on a ZSPRAY (Micromass) interface. Chromatographic separations were conducted on a reversed-phase (RP) capillary column (Pepmap C18, $75\mu$m i.d., 15 cm lenght, LC Packings) with a 200 nL/min flow. The gradient profile used consisted of a linear gradient from 95% A (H2O / 0.05% HCOOH) to 45% B ( acetonitrile / 0.05% HCOOH) in 35 min. followed by a linear gradient to 95% B in 1 min. Mass data acquisitions were piloted by MassLynx software (Micromass, Manchester, UK) using automatic switching between MS and MS/MS modes. The internal parameters of Q-TOF II were set as follows. The electrospray capillary voltage was set to 3.0 kV, the cone voltage set to 30 V, and the source temperature set to 80°C. The MS survey scan was m/z 300-1500 with a scan time of 1 s and a interscan time of 0.1s. When the intensity of a peak rose above a threshold of 8 counts, tandem mass spectra were acquired. Normalized collision energies for peptide fragmentation was set using the charge-state recognition files for +1, +2 and +3 peptides ions. The scan range for MS/MS acquisition was from m/z 50 to 1500 with a scan time of 1 s and a interscan time of 0.1s. Fragmentation was performed using argon as the collision gas and with a collision energy profile optimized for various mass ranges of precursor ions.

Mass data collected during a nanoLC-MS/MS analysis were processed and converted into a .PKL file to be submitted to the search software MASCOT (Matrix Science, London, UK).

**Statistics**

Statistical comparisons were made using the Student's two tailed $t$ test. All results are representative of at least 3 experiments and expressed as means ± SD of at least 3 replicates.

**3-Results**

Human monocyte-derived DC are classically CD1a$^+$ and CD14$^-$, opposite to monocytes. The immature DC phenotype was attested by intermediate surface expression levels of HLA-ABC and HLA-DR and by negative or low expression of CD25, CD83 and co-stimulatory molecules CD80, CD86 (Fig. 1a). Following stimulation by the bacterial-derived danger signal LPS or by the T cell-derived signal CD40-ligand, DC phenotype displayed the typical inductions of CD25, CD83 and enhancements of CD80, CD86, HLA-ABC and HLA-DR surface expressions. DC differed from monocytes in their function since they elicited allogeneic T-cell responses (Fig. 1b).

Comparative 2D gel analysis of the monocyte, immature DC and mature DC stages is shown on figures 2 and 3. The most striking and reproducible differences indicated by arrows on the figures, as pulled out from a quantitative gel image analysis, were further analyzed by mass spectrometry to determine the nature of these differentially-expressed proteins. Gels with close detection sensitivity were selected, so that the detection parameters for image analysis could be kept constant and lead to close quantitative values (see methods section). It is striking to note that there are relatively few reproducible differences, especially between immature and mature dendritic cells. This step was our initial main focus, as the monocyte-immature dendritic cell transition had been investigated before [40]. These few differences point to well known functions of dendritic cells, such as antigen presentation (HLA class II), or cytokine production (IL1). They also point out cytoskeletal remodelling (gelsolin), which is obvious when taking into account the morphological changes between monocytes, immature and mature dendritic cells. However, one of the most striking diffferences is MnSOD. This protein has been previously described as heavily induced during the monocyte-immature DC transition [40], and it is rather surprising to see that despite the high levels reached at this stage, a further induction is observed during DC maturation. As MnSOD is an oxidative stress response protein, which expression is often modulated during pro-apoptotic conditions [41] or during cancer transformation [42], this oriented us to focus our study on oxidative stress response proteins. This was further reinforced by the fact that Prx1, which is also an important oxidative stress response protein, and Trx1, which plays an important role in several cellular redox processes, were also shown to be induced at both stages, as MnSOD.
An additional benefit of the proteomics approach in the study of peroxiredoxins is that this technology allows the access to both native and oxidized forms of the proteins [43], [44]. From previously mentioned studies, we knew the positions of the various peroxiredoxins spots on the gels. We however reassessed their positions on our DC gels by mass spectrometry to further secure our identifications. Sequence coverages were always above 50% (data not shown), thereby providing safe protein identifications. The identification of the peroxiredoxins and of Mn SOD was however also secured by MS/MS analysis (see supplementary material)

The quantitative results obtained on oxidative stress proteins are summarized in Table 1.  Although they were identified on the gels as very minor spots (indicated by arrowheads), the oxidized forms of Prx1, Prx2 and Prx6 are not mentioned in the table, as they did not show any quantitative variation along DC differentiation and maturation processes. Expressions of the various proteins belonging to the Prx family are divergent and depend on the differentiation and maturation states. Some Prx decreased moderately but steadily during the differentiation and maturation process (Prx6, Prx4), while others increased steadily (Prx1), and others showed a bell-shaped expression curve (Prx2), peaking at the immature DC stage. The low but steady expression level of the mitochondrial Prx3 is surprising, as it is a mitochondrial oxidative stress response protein (as MnSOD). However, MnSOD is expressed at much higher levels and also shows a clear induction in our biological system. Finally, results similar to those obtained with CD40-ligand were also obtained upon DC maturation with LPS (data not shown). In  order to confirm those results, we used two different strategies. One is a co-migration strategy (Fig. 4). Pure Prx2 purified from human red blood cells [30] was added in increasing amounts to monocytes extracts, and the spiked extracts were separated by two-dimensional electrophoresis and submitted to image analysis. The results obtained showed a two-fold difference between the amount of Prx2 that should be added to mimic the intensity observed in DC (e.g. 20ng for immature DC) and the amount that was deduced from the quantitative image measurements (theoretically 40 ng).  This can be accounted for by postulating that half of the loaded proteins are lost in the 4-8 two-dimensional analysis, either from proteins lying outside the separation space (in pI and Mw) or from proteins lost because of poor solubility (e.g. membrane proteins). Such a quantitative yield has already been described [45]. The second strategy is western blotting, and was used for MnSOD (Fig.5). SDS-PAGE gels were used to secure against any artefact arising from 2D electrophoresis. The results clearly show an induction of the protein, as observed from 2D gels. However, the induction factors observed by blotting were clearly inferior to those observed by 2D gel electrophoresis, in accordance with the known non linear behavior of western blotting [46].

Due to their high rate of anti-oxidant enzymes, we postulated that DC are more resistant to oxidative environments than monocytes. In human advanced atherosclerosis, the atherosclerotic lesions (plaques) constitute an extensive oxidative environment. Indeed, oxLDL accumulating in these lesions, trigger the production of superoxide and peroxide by macrophages, but also directly exert a toxic activity towards different cell types, because of the lipid peroxides and hydrogen peroxide they contain. *In vitro* experiments have previously revealed that oxLDL are toxic and apoptosis-inducing for macrophages and smooth muscle cells [15], [47]. We thus studied and compared the survival capacities of both DC and monocytes exposed to high concentration of oxLDL. Cell death was measured after 24h of incubation by intracellular incorporation of propidium iodide.. The number of apoptotic monocytes following oxLDL treatment dramatically increased in the presence of oxLDL from 12% to 76% (Fig. 6a). A preliminary dose response study showed that monocytes were as potent as DC to detoxify oxLDL until a 50µg/mL concentration (30 to 35% of death). When oxLDL concentration exceeded this 50µg/mL threshold, more than 87,3% of monocytes underwent apoptosis while immature DC and mature DC were less affected (47,8 and 67,3% of death respectively) (Fig. 6b). More detailled investigations, working with 75µg/mL oxLDL doses demonstrated that immature DC and, to a lower extent, mature DC were more resistant than monocytes to the apoptosis induced by high doses of oxLDL (Fig. 6c). Thus the survival of immature DC in the oxidative stress environment generated by oxLDL in vitro is better than monocyte survival. This finding can be correlated to their respective expression levels of MnSOD and Prx.

The enhanced capacity of immature DC to survive in an oxLDL-enriched environment, compared to monocytes, as well as the presence of DC-associated T cells in atherosclerosis lesions, lead us to investigate the production of RANKL in DC/Tcell co-cultures. RANKL is the ligand of the TNF receptor family member RANK, expressed on T cells activated by anti-CD3 and anti-CD28 antibodies [48] and on memory T cells [27]. It is a crucial survival factor for DC and a chemotactic factor for monocytes [26], [28], [49]. Moreover, RANK and RANKL expressions have reported in the atherosclerotic plaques [50]. Immunofluorescent staining were used to study the expression of RANKL on T cells co-cultured with monocytes or DC. T cells co-cultured with monocytes weakly expressed RANKL and monocytes did not express RANKL themselves. In contrast, T cells co-cultured for 5 days with DC strongly expressed RANKL on their surface. Interestingly, DC also displayed RANKL expression (Fig. 7). Consequently besides MnSOD and Prx overexpressions, DC-induced RANKL expression may also be a key parameter accounting for a role of DC in atherosclerosis pathogenesis.

## 4-Discussion

Because of their outstanding interest in immunology, gene expression in DC has been studied in details, both at the monocyte-immature DC transition [40], [51], and at the immature-mature DC transition [52], [53]. Proteomic study complementing the latter transcriptomic studies on DC maturation was required. We focussed our study on a major protein induction observed in proteomics: MnSOD reaches (by combining the two major MnSOD spots) 0.7% of the gel-displayed proteins at the mature DC stage. This important amount oriented us to study the oxidative stress response pathways. MnSOD is up-regulated by a variety of pro-inflammatory mediators, such as TNF-$\alpha$, LPS, IL-1$\beta$ and IFN-$\gamma$ [54], [55]. Therefore, the observed induction of MnSOD protein can be correlated with its well-documented role as an anti-apoptotic protein [41], [56]. The main activity of MnSOD enzyme is to reduce anion superoxides into hydrogen peroxides. Together with glutathione peroxidase and catalase, Prx enzymes play an important role in eliminating peroxides [57], which are produced by numerous pathways, including the dismutation of superoxide. For example, Prx1 and Prx2 are involved in the removal of $H_2O_2$ in thyroid cells and can protect these cells from undergoing apoptosis [58]. Only a moderate increase of DC apoptosis was observed in the presence of $H_2O_2$, [59] probably due to their high Prx expression. Strikingly, $H_2O_2$ rather seems to be an activating signal for DC, as it stimulates their production of IL-8 and TNF-$\alpha$ in a dose-dependent manner [10].

The high amount of antioxidant enzymes awards DC the ability to survive in highly oxidant environment. This also correlates with the anti-apoptotic effects observed upon DC maturation [60]. Recent studies demonstrated that oxLDL promote DC differentiation from monocytes [25], whereas oxLDL induce apoptosis of human macrophages, a main feature in the first steps of atherogenesis [61]. As oxLDL produce both organic peroxides and hydrogen peroxide [14], it was interesting to study the changes in Prx expression during DC differentiation and maturation. In this case, the proteomic study is especially well-suited, as the native and oxidatively-inactivated forms of Prx can be separated on 2D gels [43], [44]. In the case of DC differentiation and maturation, we did not notice any changes in these oxidized forms, which were present at very low amounts. However, interesting quantitative findings could be made on the normal forms of Prx. First of all, the relative amounts of the various Prx are very different. Mitochondrial Prx3 was always expressed at low levels. However, among the cytosolic Prx (Prx1, 2 and 6), Prx2 is expressed at much lower level than Prx1 or Prx6 in monocytes (350 versus above 2000 ppm), while this is not the case in Jurkat cells (where Prx6 is almost undetectable [43]) or in HeLa cells, where Prx2 is expressed at much higher levels [44]. Prx2 counteracts the NF-κB pathway [57], required for DC maturation [60]. It is therefore not surprising to find low levels of Prx2 in mature DC and, more generally, low levels of Prx2 in the monocyte-DC lineage compared to other haematopoietic lineages [62].

It is generally believed that all mammalian cytosolic peroxiredoxins have similar substrate specificity and are thus able to destroy both hydrogen peroxide and organic peroxides [63]. In terms of peroxide destruction, the total cytosolic peroxiredoxin amount (i.e. Prx1 + Prx2 + Prx6) increases from 4800 ppm at the monocyte stage to 7500 ppm at the immature DC stage and to 9300 ppm at the mature DC. Although Prx6 also exhibit other different actions, which may alter this simplistic scheme, our results suggests that the Prx defence line is induced steadily during the DC differentiation and maturation. This overall increase, however, masks divergent quantitative changes in the cytosolic peroxiredoxins upon DC differentiation and maturation, which is the most surprising part of our data. Prx1 continuously increases during this process (close to 3 fold) while Prx6 continuously decreases (2 fold) and Prx2 peaks at the immature DC stage.

An interesting trend is provided by the dual function of Prx6, which is at the same time a peroxiredoxin and a phospholipase A2 (PLA2) [64]. Thus, a decrease in Prx6 amount also means a decrease in PLA2 activity. Easily oxidized polyunsaturated fatty acids are often found at the 2-position in phospholipids, and are therefore liberated by PLA2 activity. Inhibition of PLA2 can be of physiological interest since it prevents the rise of lysophosphatidylcholine levels and diminishes the death-inducing effects of oxLDL on monocytes [65]. Apart from the context of oxidized lipids, it must be kept in mind that Prx6 is also an activator of NADPH oxidase [66], which is active in DC and especially mature ones [67]. All these factors may explain the reorientation from Prx6 to Prx1, occurring during DC differentiation and maturation.

In a more general frame, our data concerning monocytes and DC survival in an oxLDL-enriched stress environment can not be directly correlated to the global cellular expression levels of anti-oxidant enzymes Prx. Indeed, mature DC, which have the highest contents in Prx, are more sensitive to oxLDL-induced apoptosis than immature DC. Cell survival to oxLDL-induced death correlates rather well with the cytosolic amount of Prx-2. This is rather surprising, as the various mammalian Prx are known to be able to reduce the same scope of peroxides *in vitro* [63]. However, it must be kept in mind that Prx interact with multiple and different partners [68]. These other interactions may alter their operational efficiency, either by altering their catalytic efficiency or by segregating some types of Prx away from the peroxide substrates generated by oxLDL, expecially lipid hydroperoxides. This may explain why reistance to oxLDL correlates with the expression of one particular Prx and not with the total Prx amount.

Concerning the role of DC in atherosclerosis, it is now accepted that mature DC, which are known to be present in advanced plaques [69], contribute to plaque destabilization especially through T cell activation and CD40–CD40L interactions play a key role in atherosclerosis progression [70]. Besides being implicated in the induction of local immune or auto-immune responses, our results indicate, that when interacting with T cells, DC produce the cytokine RANKL themselves and induce its production by T cells. RANKL, also called osteoprotegerin-ligand (OPGL) or TNF-related activation-induced cytokine (TRANCE), is a chemotactic factor for monocytes, together with MCP-1 produced by endothelial cells, following oxLDL stimulation in atherosclerosis. RANKL expression has been detected in advanced calcified lesions of atherosclerosis [50]. Thus DC and T cell-derived RANKL might also enhance the recruitment of monocytes from blood towards the atherosclerosis lesions.

RANKL also plays important roles in DC homeostasis. RANKL-RANK interaction has been shown to sustain DC survival, by inducing the anti-apototic gene Bcl-xL [28]. Skin CD1a$^+$ DC express RANK but lack RANKL and are short-lived. However, they can be rescued from cell death, either by recombinant soluble RANKL or by RANKL+ DC generated in vitro from CD34$^+$ progenitors [28]. In addition to enhancing DC survival, RANKL induces the expression of proinflammatory cytokines (IL-6, IL-1) and T cell growth and differentiation factors (IL-12, IL-15) by DC in vitro [27]. RANKL also provides co-stimulation required for efficient CD4+ T cell priming during viral infection in the absence of CD40L/CD40 [49], [71]. These data further suggest that, besides recruiting more monocytes to the atherogenic plaque, RANKL produced locally may also increase DC lifespan in the plaque and amplify their functions, then contributing to the progression of atherosclerosis.

In conclusion, DC recruitment, maturation and survival may be critical factors for the progression of atherosclerosis: (i) DC recruitment into the vascular wall is increased by atherogenic stimuli such as oxLDL, TNF-α, and hypoxia [72], (ii) DC maturation is induced by different atherogenic stimuli such as superoxide, oxLDL, lysophosphatidylcholine, nicotine, angiotensin II, atrial natriuretic peptide, and TNF-α [7-10], [73], [74] (iii) DC survival in the atherosclerotic plaque environment is a third important factor enabling these professional antigen presenting cells to induce a more important antigen-specific T cell activation. The high anti-oxidant enzyme expression levels, the better resistance to oxLDL-induced apoptosis and the production of RANKL upon DC/T cell interactions, that we highlighted in this work are three elements which further implicate DC in the pathogenesis of atherosclerosis and emphasize the role they may play in the amplification of the chronic inflammation, together with hypercholesterolemia and oxidative stress. This role of DC in atherosclerosis progression was recently further supported by observations indicating that oxLDL and platelet-activating factor contained in the lesions locally activate DC, but inhibit their migration to draining lymph nodes [75]. This activated-DC sequestration, in addition to enhanced DC survival, may further aggravate the local inflammation.

## Acknowledgements

TR wants to acknowledge personal support from the CNRS. This work was supported by grants from INSERM, UCBL, CNRS, emergence project of Rhone-Alpes Region, ARC 4800 and 3637, Ligue contre le Cancer Ardèche, Drôme and Rhône and MENRT ACI 8BC05H.

**Figure legends**

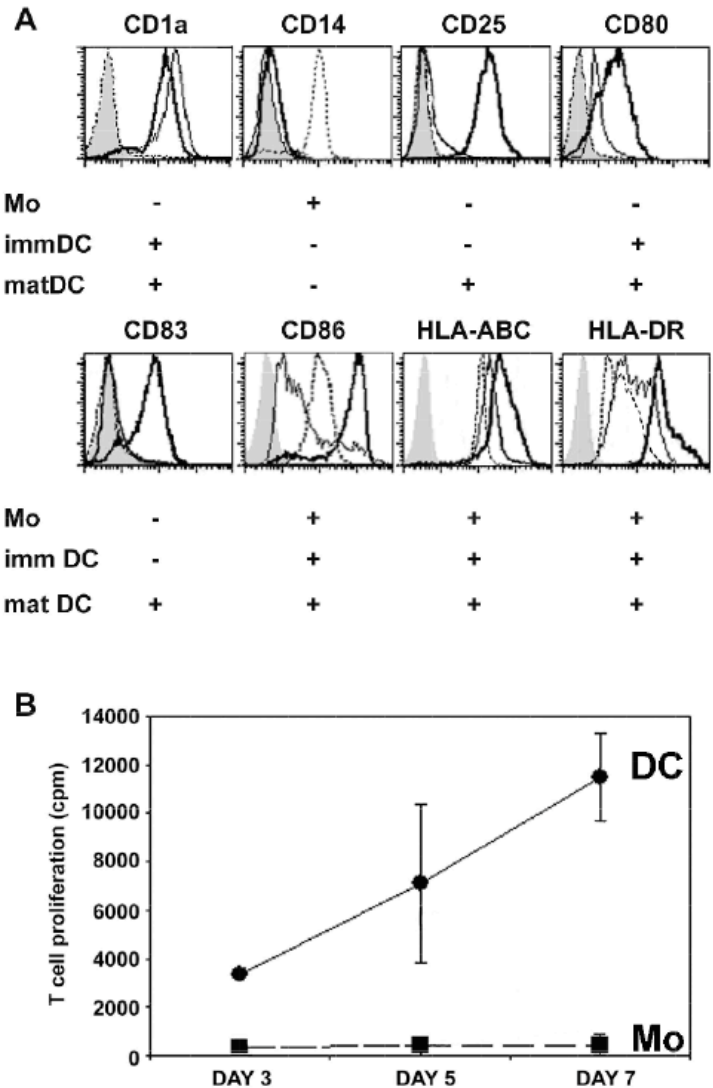

**Figure 1: Phenotype and function of human monocytes and DC.**
(A) Phenotype analysis of peripheral blood monocytes (dashed line), monocyte-derived immature DC (thick line) and CD40L-activated mature DC (bold line) by cytofluorimetry. Cells were stained with antibodies against CD1a, CD14, CD25, CD80, CD83, CD86, HLA-ABC and HLA-DR or related isotype control antibodies (gray histogram). Data are representative of more than 10 experiments.
(B) T cell proliferation measured by thymidine incorporation, at indicated time points, in coculture of allogeneic T cells with either DC (circle) or monocytes (square). DC: T cell ratio was 1:100 cells. Triplicate experiment is representative of more than 10 experiments.

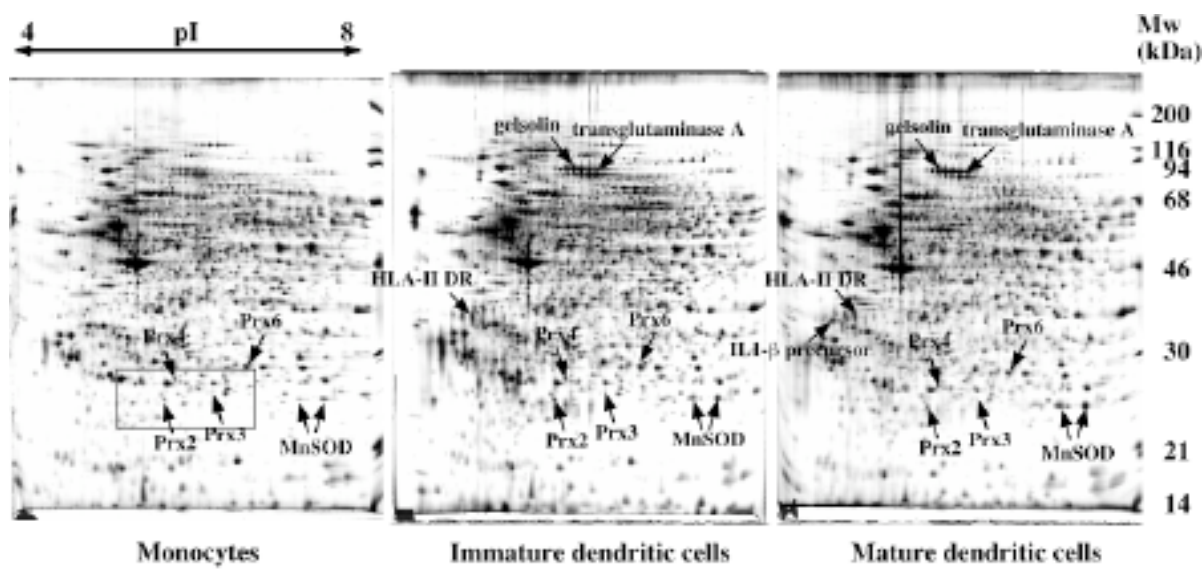

**Figure 2: 2D electrophoresis of whole cell extracts (acidic and neutral proteins)**
Whole cell extracts (120 $\mu$g protein) prepared from monocytes, immature and mature dendritic cells were analysed by 2D electrophoresis. The pH gradient in the first dimension ranged from 4 to 8, thereby separating the acidic and neutral cellular proteins only. The second dimension was a 10% gel, using a pH 8 gel and the taurine system. The oxidative stress response proteins identified on the gels are shown by arrows. The boxed zone is the one shown in figure 4

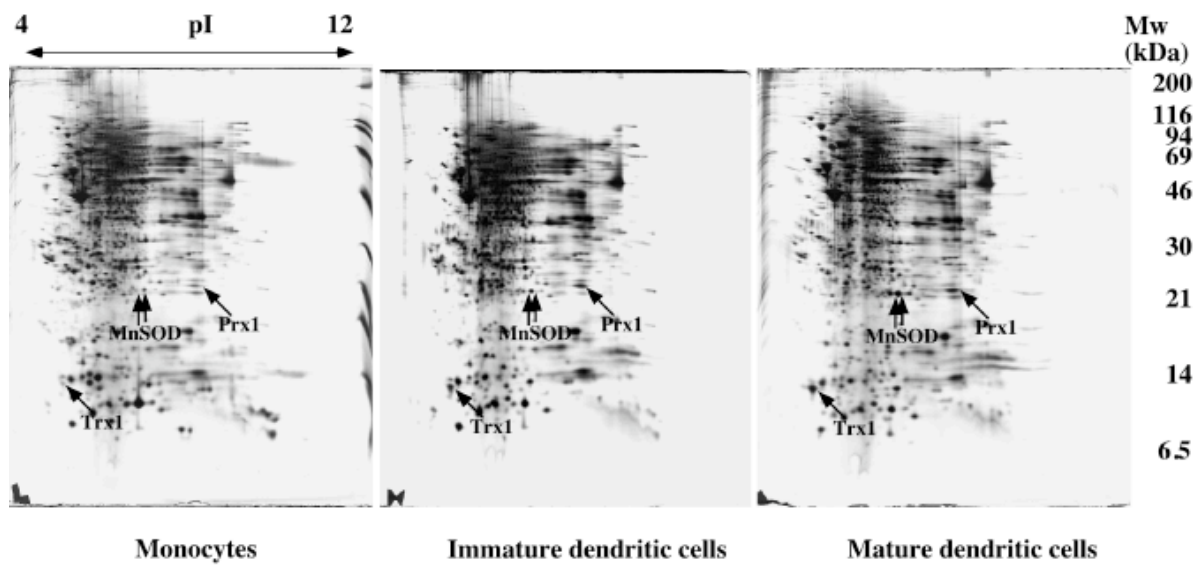

**Figure 3: 2D electrophoresis of whole cell extracts (basic proteins)**
The same analysis as in figure 2 was performed, but a pH 4-12 gradient was used in the first dimension. The second dimension was a 11% gel, using a pH 8.2 gel and the taurine system. This allowed to analyze the basic proteins. Silver staining with ammoniacal silver.

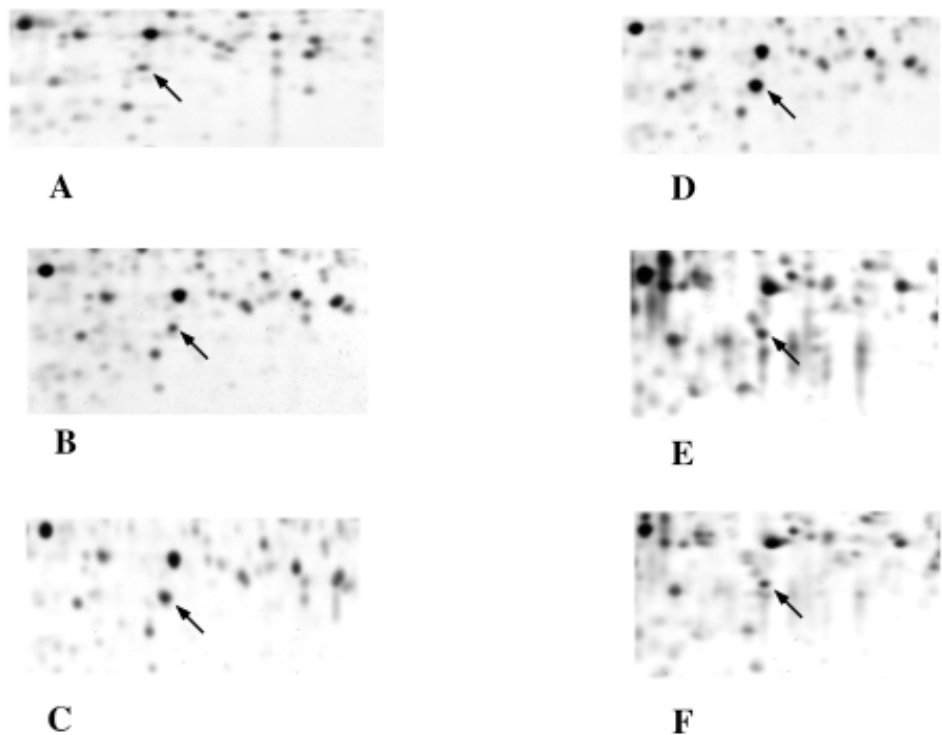

**Figure 4: Comigration of Prx2 with monocyte extracts**
Only Prx2 zone is shown in this figure. Increasing amounts of chromatographically purified Prx2 [30] were added to a monocyte extract, and the spiked extracts were separated by two dimensional electrophoresis using linear 4-8 pH gradient. A) starting monocyte extract. B) monocyte + 10ng Prx2. C) monocyte + 20ng Prx2. D) monocyte + 50ng Prx2. E) immature dendritic cells. F) mature dendritic cells. Arrows points out Prx2 spots in the different conditions.

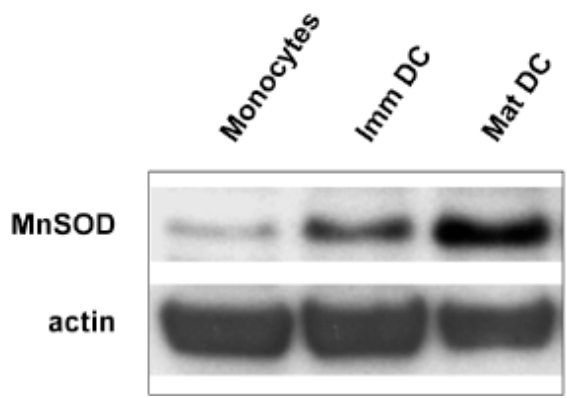

**Figure 5: Western blotting of MnSOD**
Western blot analysis of MnSOD expression in monocytes, immature monocyte-derived DC (Imm DC) and LPS- matured monocyte-derived DC (Mat DC). Immunoblots analysis was performed with anti-MnSOD antibody (#06-984, Cell Signaling) and anti-actin (A-2066, Sigma Aldrich) antibody as loading control. Data are representative of two experiments.

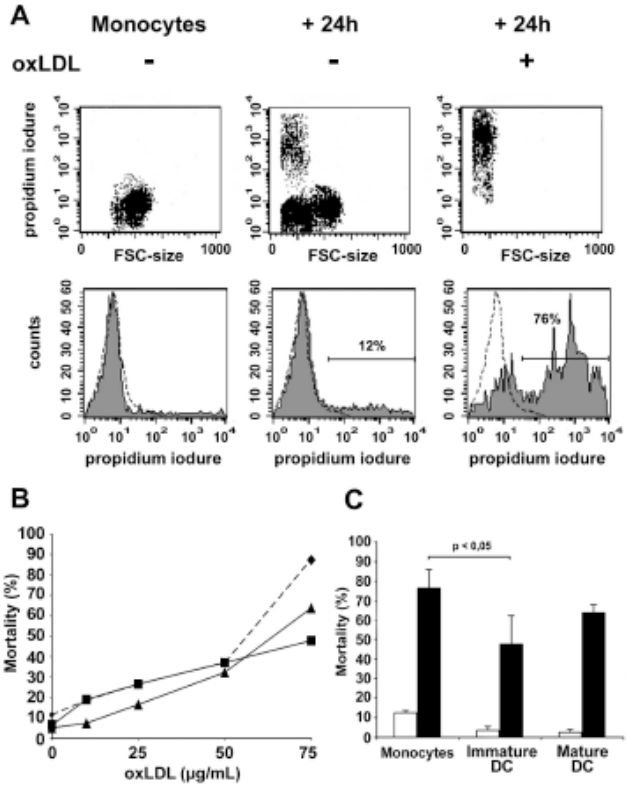

**Figure 6: Apoptosis in human monocytes and DC untreated or treated with oxLDL.**
Apoptosis was quantified by cytofluorimetry after incorporation of propidium iodide.
(A) Monocytes were cultured 24h with or without 75 _g/mL OxLDL.
(B) Monocytes (diamonds), immature monocyte-derived DC (squares) and CD40L-activated mature DC was cultured 24h with increasing dose of OxLDL.
(C) Mortality of monocytes, immature monocyte-derived DC and CD40L-activated mature DC was measured after 24h culture in presence (black bars) or in absence (white bars) of 75 _g/mL OxLDL. Means and SD of three experiments.

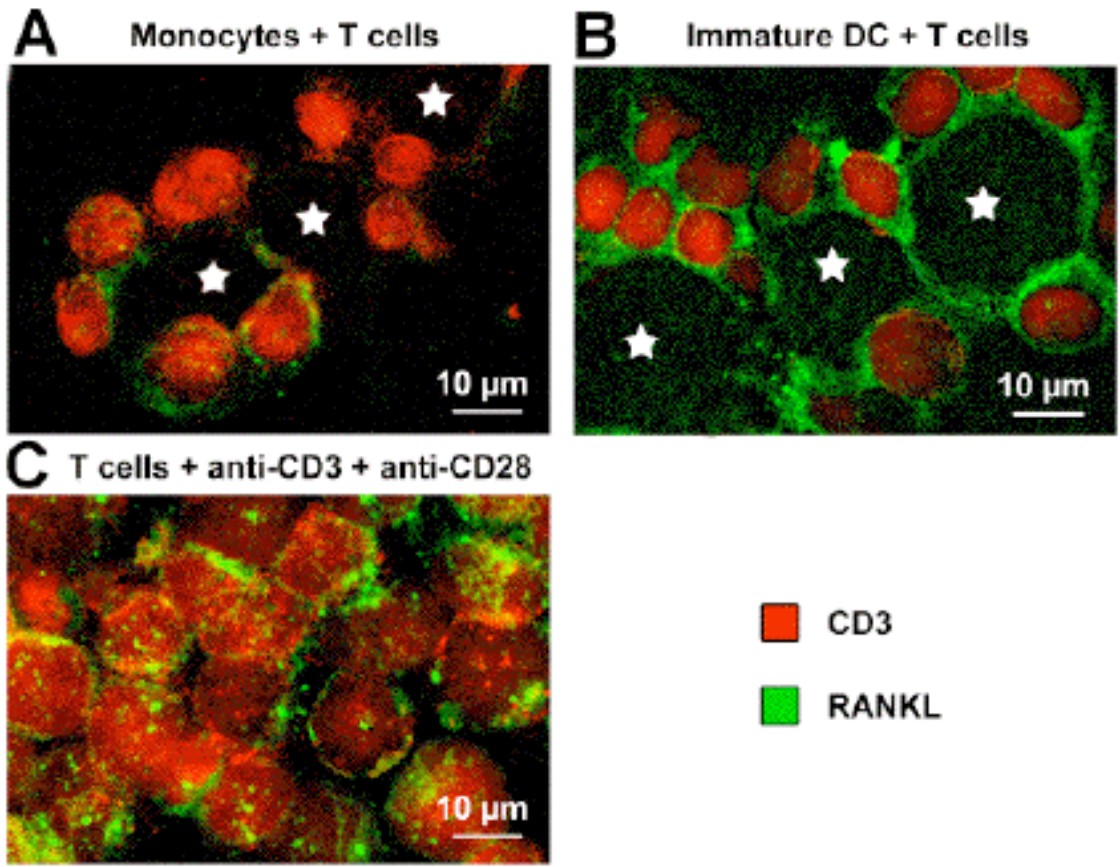

**Figure 7: RANKL expression in cocultures of T cells with either monocytes or DC.**
RANKL (green) and CD3 (red) stainings of T cells, after 5 days of culture with (A) monocytes, (B) immature DC and (C) anti-CD3 plus anti-CD28 antibodies. Stars indicate monocytes (A) or DC (B). T cells were distinguished from other cells based on CD3 expression and Hoechst nuclei staining (not shown).

Table 1: quantitative measurements of the oxidative stress response proteins [a, b]

|                      | Monocytes      | Immature DC     | CD40L-activated DC |
|----------------------|----------------|-----------------|--------------------|
| Mn SOD, basic spot*  | 1260±80 ppm    | 2010±120 ppm    | 4550±200 ppm       |
| Mn SOD, acidic spot* | 250±30 ppm     | 1020±80 ppm     | 2450±180 ppm       |
| Prx1†                | 2150±100 ppm   | 4960±240 ppm    | 7440±460 ppm       |
| Prx2                 | 350±30 ppm     | 680±40 ppm      | 520±40 ppm         |
| Prx3                 | 40±15 ppm      | 50±15 ppm       | 30±10 ppm          |
| Prx4                 | 850±80 ppm     | 560±50 ppm      | 520±40 ppm         |
| Prx6                 | 2240±220 ppm   | 1810±140 ppm    | 1350±110 ppm       |

The values are expressed in ppm of the total spot volume, as calculated by the Melanie software. Due to the precision of the measurement (typically± 10%, but indicated on the table) they have been rounded (to the closest decennial)

Study of the variations in expression of the antioxidant proteins between the various cellular stages showed significant variations ($p<0.05$) except for Prx3 (no variation), Prx4 between immature and mature DC ($p<0.5$), and Prx6 between monocytes and immature DC ($p<0.1$)

*: Mn SOD can be measured on pH 4-12 and on pH 4-8 gradients. The values are those obtained on pH 4-8 gradients
†: Prx 1 can be measured only on pH 4-12 gradients. All other Prx are measured on pH 4-8 gradients